\documentclass[%
 reprint,
 amsmath,amssymb,
 aps,
]{revtex4-2}

\usepackage[dvipdfmx]{graphicx,color}	
\usepackage{dcolumn}
\usepackage{bm}
\usepackage{braket}
\usepackage[mathlines]{lineno}
\usepackage[hyperindex,breaklinks]{hyperref}

\begin{document}

\preprint{APS/123-QED}

\title{Gate controlled unitary operation on flying spin qubits in quantum Hall edge states}

\author{Takase Shimizu}
\author{Taketomo Nakamura}%
\author{Yoshiaki Hashimoto}
\author{Akira Endo}
\author{Shingo Katsumoto}
\affiliation{%
 Institute for Solid State Physics, The University of Tokyo, 5-1-5 Kashiwanoha, Kashiwa, Chiba 277-8581, Japan
}%


\begin{abstract}
Spin and orbital freedoms of electrons traveling on spin-resolved quantum Hall edge states (quantum Hall ferromagnets) are maximally entangled.
The unitary operations on these two freedoms are hence equivalent, which means one can manipulate the spins
with non-magnetic methods through the orbitals.
If one takes the quantization axis of spins along the magnetization axis, the zenith angle is determined by the partition rate of spin-separated edges
while the azimuth angle is defined as the phase difference between the edges.
Utilizing these properties, we have realized electrically controlled unitary operation on the electron spins on the quantum Hall ferromagnets.
The zenith angle of the spin was controlled through the radius of gyration at a corner by means of applying voltage to a thin gate placed at one edge.
The subsequent rotation in the azimuth angle was controlled via the distance between the edge channels also by a gate voltage.
The combination of the two operations constitutes the first systematic electric operation on spins in the quantum Hall edge channels.
\end{abstract}

\maketitle


\section{\label{sec:level1}Introduction}
Electron spins in semiconductor nanostructures like quantum dots, have been expected to serve as qubits for practical quantum computation\cite{PhysRevA.57.120}.
The flying qubit (FQ) scheme, which is the usage of traveling electron spins as qubits, is not only indispensable for long-distance entanglement\cite{PhysRevB.93.075301,Fujita2017}
between localized arrays of qubits, but also usable for unitary operation on qubits\cite{Yamamoto2012}.
The flying spin qubits (FSQs) for electrons considered and tested so far were mostly based on quantum wires with Rashba and Dresselhaus spin-orbit interactions (SOIs)\cite{PhysRevB.71.033309,Gong2007,NadjPerge2010}, which work as a magnetic field effective only on spins\cite{rolandwinkler2003}.
A convenient way to describe unitary transformations on an FQ is obtained by viewing the traveling electron from the coordinate fixed to the center of
the electron wave packet (center coordinate).
Then we can introduce a spatially local Hamiltonian $H_{\rm loc}$ to describe the dynamics other than the translational motion. That is, travel of wave packet through a quantum wire with spatial potential modulation
can be viewed as a process, in which the local Hamiltonian evolves with time\cite{Bttiker2000,PhysRevLett.87.256602}, {\it i.e.}
the effective Schr\"odinger equation for wave packet $\ket{\Phi}$ on the center coordinate is written as
\begin{equation}
i\hbar \frac{\partial \ket{\Phi}}{\partial t}=H_{\rm loc}(t)\ket{\Phi}.
\label{eq_wave_packet_travel}
\end{equation}
In this picture, the SOI term in the Hamiltonian also varies with time, which can cause unitary transformation of the spin freedom in $\ket{\Phi}$
through non-adiabatic transition.

Here we would like to show there is another type of SOI without a term that explicitly contains $\hat{p}$ (momentum operator) and $\hat{s}$ (spin operator) in $H_{\rm loc}$.
Generally, an interaction term in a Hamiltonian introduces quantum entanglement\cite{Schrdinger1935} between initially unrelated subsystems.
On the other hand, when we prepare an initial state with finite quantum entanglement,
the interaction appears between the subsystems without any interaction term in the hamiltonian\cite{QuantumEntanglementInteractionAndTheClassicalLimit}, as is well known as the EPR paradox\cite{PhysRev.47.777}.

For example, let us consider a wave packet $\ket{\Phi}$ with the Stern-Gerlach type entanglement\cite{Gerlach1922} between orbital $\{\ket{\xi},\ket{\eta}\}$ and spin $\{\ket{\uparrow},\ket{\downarrow}\}$ as
 \begin{align}
  \ket{\Phi}
  =
  \ket{\xi}\left(\cos \frac{\theta}{2} \ket{\uparrow}\right)
  +
\ket{\eta}\left( e^{i\phi} \sin \frac{\theta}{2} \ket{\downarrow}\right),\label{eq_fq_scheme}\\
  (0\leq\phi<2\pi, 0\leq\theta\leq\pi),\nonumber
 \end{align}
where $\braket{\xi|\xi}=\braket{\eta|\eta}=1$, $\braket{\xi|\eta}=0$,
$\phi$ and $\theta$ are the azimuth and zenith angles of the spin, respectively.
We assume $H_{\rm loc}(t)$ has no operator on $\{\ket{\uparrow},\ket{\downarrow}\}$, thus no explicit interaction term for a certain period,
in which the orbital part evolves to $e^{i \chi}(\ket{\xi},e^{i\varphi}\ket{\eta})$.
Here $\chi$ is the phase developed with the travel of wave packet while
$\varphi$, the phase difference due to path difference etc., can be absorbed into the azimuth angle of the spin.
The process hence can be viewed as a translational motion with spin precession, 
which is nothing but a phenomenon with an effective magnetic field by an SOI\cite{Iwasaki2017}.
The equivalence of the phenomena with a quite different appearance
reflects the inseparability of systems with a maximal entanglement.
This idea also suggests a possibility to manipulate the electron spins through orbital motion, and
the architecture is theoretically proposed in Ref.\cite{PhysRevB.77.155320}.

The quantum coherence length of the one-dimensional channel for the FQ propagation should be long enough to preserve quantum information.
In solids, the longest coherence lengths were reported for quantum Hall edge channels (QHECs)\cite{Machida1998,PhysRevLett.100.126802},
which are thus strong candidates for FQ channels.
The QHEC is known to show spin separation at comparatively low magnetic fields with the aid of exchange interaction\cite{PhysRevB.37.1294},
and is transformed into ferromagnetic phases\cite{PhysRevB.31.6228,Piazza1999}.
In such a ferromagnetic regime, the entanglement of spin and edge channel occurs,
naturally preparing realization of the scheme in \eqref{eq_fq_scheme}.
A preliminary experiment on such precession control was reported by Nakajima {\it et al.}\cite{Nakajima2012, Nakajima2013}.
For a zenith angle tuning, a controlled splitting of wave packet into two channels is required.
Such experiments have been reported by Deviatov {\it et al.} with use of current imbalance\cite{Deviatov_2012, PhysRevB.77.161302, PhysRevB.84.235313}, and by Karmakar {\it et al.} with use of periodic magnetic gates\cite{PhysRevLett.107.236804,PhysRevB.92.195303}.

In this article, we present experimental results on unitary operations of FSQ in spin-polarized QHECs
in the above scheme with electrostatic gates.
In the view of eq.\eqref{eq_fq_scheme}, a rotation in the zenith angle corresponds to a tunneling between
spin-polarized QHECs.
It is shown that this can be achieved by a sharp bending of the edge line.
At such a corner, an angular momentum in the edge orbital emerges, and the Landau-Zener type tunneling brings about a rotation in
the zenith angle.
The rate of the Landau-Zener tunneling depends on the sharpness of the corner.
With the addition of a thin gate to change the sharpness, we show that the zenith angle can be controlled more simply.

\section{Experimental Method}
\begin{figure}[t]
  \includegraphics[width=1\linewidth]{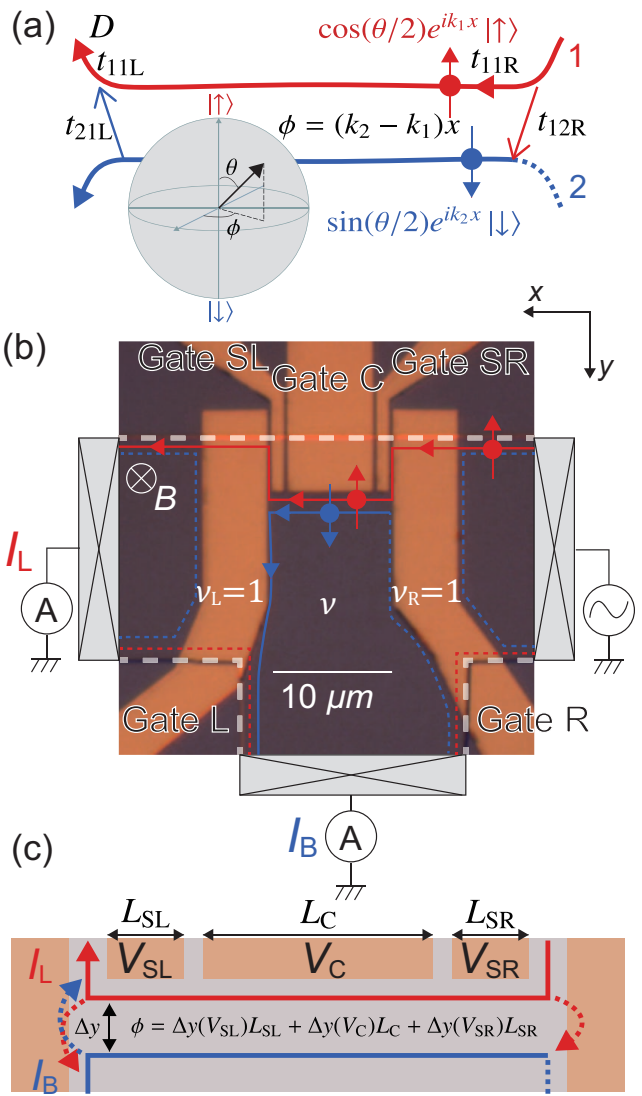}
  \caption{ (a) Schematic diagram of the ``quantum circuit" for electron wave packets (red and blue circles with arrows indicating spins),
with an illustration of Bloch sphere description of an FSQ.
(b) Optical micrograph of the sample with the external circuit illustration. 
The orange regions are metallic gates, three of which are annotated. 
The 2DES substrate is trimmed at the white dashed lines.
(c) illustrates a hybridization of (a) and (b) around the lower ends of gate SL, C, and SR. }
  \label{fig-expl}
 \end{figure}

Figures \ref{fig-expl}(a)-(c) describe the experimental setup in three different ways
for a two-dimensional electron system (2DES) in the spin-split quantum Hall regime.
For simplicity, the filling factor $\nu$ is chosen to 2 in the figure, though the region of $\nu=4$ was mostly used in the present experiment.
Figure \ref{fig-expl}(a) is a schematic of wave-propagation paths, (b) shows the gate-electrode configuration for realizing them,
and (c) illustrates a blowup of down edges of side-gates (SL, SR) and center gate (C) along with the propagation paths in (a).
The sample edges have two QHECs for $\nu=2$, and here we call them channel 1 and channel 2, in which spins are locked to $\uparrow$ and $\downarrow$ respectively.
Then we can write wave packet on them as $\ket{1}\ket{\uparrow}$ and $\ket{2}\ket{\downarrow}$, where
$\ket{1}$ and $\ket{2}$ are normalized wavefunctions of the orbital part.

Let us trace a wave packet which is emitted from the right electrode.
Beneath the Gate L and Gate R, the filling factors $\nu_{\rm L}$ and $\nu_{\rm R}$ are tuned to 1, and only channel 1 goes through them\cite{Hashisaka2017,Hashisaka2018}.
Hence, the incidence wave packet in Fig.\ref{fig-expl}(a) can be written as $\ket{\psi_1}\ket{\uparrow}$.
Channels 1 and 2 meet at the lower right corner edge of gate SR, where
a partial transfer occurs through a local SOI due to the orbits wraparound the sharp corner.
This scattering process is written as
\begin{equation}
\ket{1}\ket{\uparrow}\rightarrow \ket{\Phi}_{\rm SR}= t_{\rm 11R}\ket{1}\ket{\uparrow}  + t_{\rm 12R}\ket{2}\ket{\downarrow},
\label{eq_Phi_SR}
\end{equation}
where $t_{ij{\rm R}}$ are complex transmission coefficients of the processes $\ket{i}\rightarrow\ket{j}$ at the right corner
satisfying the unitary condition $|t_{\rm 11R}|^2+|t_{\rm 12R}|^2=1$ due to the perfect chirality of QHEC.
Hence they can be written as $t_{\rm 11R}=\cos \theta/2$ and $t_{\rm 12R}=e^{i\phi_0}\sin\theta/2$,
where $\theta$ reflects the amplitude ratio of partial waves, and $\phi_0$ is the phase difference between $t_{\rm 1R}$ and $t_{\rm 2R}$.
Thus $\ket{\Phi}_{\rm SR}$ is prepared in the form of $\ket{\Phi}$ in \eqref{eq_fq_scheme}.
 
In QHEC, the orbital part of the wave function in the single-electron picture is written
as a quasi-one-dimensional plane wave in real space representation\cite{Yoshioka2002},
 \begin{subequations}
 \begin{align}
  \psi_i({\bm r})
\propto&
 \exp(ik_i x)\exp\left[-\frac{(y-y_i)^2}{2l^2}\right],
 \label{eq_orbit_wavef}\\
 &y_i=-l^2k_i, \label{eq_y-k_relation}
 \end{align}
\label{eq_single_electron_edge}
\end{subequations}
where $l(=\sqrt{\hbar/eB},~B$ being the magnetic field) is the magnetic length, $i$ is the channel number,
$x$ axis direction is taken as along the one-dimensional channel, and $y_i$ is the guiding center position.
In the edge states, $\psi_i$ accommodates propagating wave packet $\ket{i}$.
$\ket{i}$ travels on $\psi_i$ alongside the down edges of gate SR, C, and SL, gaining the kinetic phase.
At the end of the travel over total length $L$,
the difference in the acquired kinetic phase or
the azimuth angle rotation of the spin is
\begin{subequations}
 \begin{align}
\phi&=(k_1-k_2)L, \label{eq_phi}\\
&=\dfrac{(y_2-y_1)L}{l^2}=2\pi\dfrac{\varDelta y LB}{h/e}.\label{eq_phi_AB}
 \end{align}
\end{subequations}
Since $\varDelta y LB/(h/e)$ is the magnetic flux piercing the area between the two paths measured in the unit of flux quantum ($h/e$), the difference in kinetic phase acquired in the travel over the two channels is equal to the Aharonov-Bohm (AB) phase by the magnetic field.
Then we can tune $\phi$ by both the magnetic field and the voltage to gate C, which varies the interval $\varDelta y$ between the edge states.
This single-electron picture needs correction due to the screening effect, as will be discussed in the analysis section though the above can still be applied
to real experiments with some modifications.

At the left corner of Gate SL, channel 1 with $\uparrow$ goes up to go beneath the region of $\nu_{\rm L}=1$ while
channel 2 with $\downarrow$ goes down to turn around the region.
Because both channels change their directions abruptly,
a crossing transition between them by a local SOI occurs at the corner as illustrated in Figs.\ref{fig-expl}(a), (c).
In this 2-in-2-out vertex, the partition ratio is affected by the phase $\phi-\phi_0$, and the traverse across the sample
ends up at drain L or drain B with the probabilities determined by the ratio.
Hence the partition ratio can be measured as the ratio of the current through L ($I_{\rm L}$) to the whole current ($I_{\rm L}+I_{\rm R}$), {\it i.e.} current distribution ratio $D\equiv I_{\rm L}/(I_{\rm L}+I_{\rm R})$.
In a simple model of 2-in-2-out vertex described in Sec.\ref{sec_zenith}, $D$ is written as,
 \begin{align}
  D  =  C_0 + C_1\sin\theta\cos(\phi+\varDelta \varphi),
 \label{eq_D_C01}
 \end{align}
where $\varDelta\varphi$ represents the phase shift associated with inter edge scatterings at the two corners including $-\phi_0$.
Equation \eqref{eq_D_C01} is just like the simplest Young's double-slit approximation of an AB interferometer
because of the chirality or the broken time-reversal symmetry of the channels and the multi-terminal configuration\cite{PhysRevB.66.115311}.
The partition ratio of the input affects the visibility giving the $\theta$-dependence.

A two-dimensional electron system(2DES) with the electron density of $4.4\times10^{11}~\rm{cm^{-2}}$ and the mobility of $86~\rm{m^2/Vs}$ in an ${\rm Al}_x{\rm Ga}_{1-x}{\rm As}$/GaAs  ($x$=0.265) single-heterostructure was used as the base system for the sample. 
The structure of the wafer was (from the front surface) 
5 nm Si-doped GaAs cap layer, 40nm Si-doped ($N_{\rm Si} = 
2 \times 10^{18} \rm{cm}^{-3}$) ${\rm Al}_x{\rm Ga}_{1-x}{\rm As}$ layer, 15nm undoped ${\rm Al}_x{\rm Ga}_{1-x}{\rm As}$ spacer layer, and 800 nm GaAs layer with 2DES residing near the interface to the upper layer.
The terminal and Au/Ti gate configuration are shown in Fig. \ref{fig-expl} (b). We cooled the sample down to $20~{\rm mK}$ and applied perpendicular magnetic field $B$ up to 9 T, 
at which the 2DES is in the quantum Hall state with the filling factor of $\nu=2$. 

A constant ac voltage of typically $33~\mu$V-rms (except for measurements in Fig.\ref{fig:SideGateDiff}) at $170 {\rm Hz}$ was applied to the right-side contact, and
the current was measured at Drain L and Drain B with an I-V amplifier by standard lock-in measurement.
A representative difference of contact and cable resistance between Drain L and B was less than about 2\%.
Therefore, $D$ is nearly equal to the transmission probability
from Source to Drain L.

The voltage on Gate C ($V_{\rm C}$) modifies the potential gradient in $y$-direction alongside Gate C, and thus the length between neighboring edge states $\varDelta y$,
which leads to the modulation of $\phi$\cite{Nakajima2012, Nakajima2013}.

\section{Rotation in azimuth angle}
 \begin{figure}[t]
  \includegraphics[width=1\linewidth]{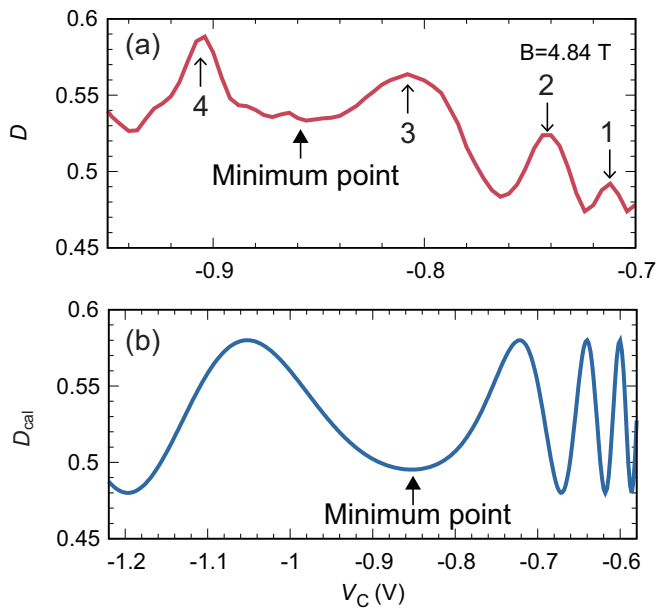}
  \caption{(a) $V_{\rm C}$ dependence of the current distribution ratio $D$ at $B=4.84~{\rm T}$.
Side gate voltages were set to $V_{\rm SR}=V_{\rm SL}=-0.6~{\rm V}$, where 2DES under the gates were totally depleted.
(b) Calculation example of $D$ as a function of $V_{\rm C}$ based on the model of eq.\eqref{eq_dx}.
$d$=66~nm from one-dimensional Poisson-Schr\"odinger calculation on the layered structure of the sample (See the supplemental material).
Other parameters are: $g =-0.6$\cite{PhysRevB.37.1294}; $\epsilon$=12.35\cite{doi:10.1063/1.88755};
$C_1\sin(\theta)$=0.05; $C_0$=0.53; $\Delta \phi=0$; $L_{\rm C}=7.5~{\rm \mu m}$; $L_{\rm SL}=L_{\rm SR}=1.25~{\rm \mu m}$; the offset in the gate voltage $V_{\rm C}$, $V_{\rm SL}$, and $V_{\rm SR}$ due to the contact built-in potential is $V_{\rm offset}=-0.05~{\rm V}$.
See the text for $n(y)$.
}
\label{fig:phi-modulate}
\end{figure}
Figure \ref{fig:phi-modulate}(a) shows $V_{\rm C}$ dependence of $D$ measured at $B=4.75~{\rm T}$, which corresponds to $\nu=4$ in the non-gated region.
The filling factors underneath the Gate L and the Gate R were kept to 1, and then the channel indices $i=$1, 2 are under our consideration (pinch-off conductance traces for Gate L are in the supplemental material).
Side gate voltages were set to $V_{\rm SR}=V_{\rm SL}=-0.6~{\rm V}$, where 2DES under the gates were completely depleted.
The measured $D$ shows an oscillation against $V_{\rm C}$ in the range from $-0.7~{\rm V}$ to $-0.98~{\rm V}$,
where four peaks are observable as indicated by arrows.
The oscillation period increased with negative $V_{\rm C}$.
The region between peaks 3 and 4 is especially wide, and the line shape shows rewinding of oscillation.
We have tested several other samples with essentially the same gate configuration, and such behavior was commonly observed.

To check the above scenario of phase modulation (or rotation in $\phi$ equivalently),
we need to know how $\varDelta y$ in \eqref{eq_phi_AB} depends on $V_{\rm C}$ taking the electric screening effect into account.
In the single-electron picture of eq.\eqref{eq_single_electron_edge}, the one-dimensional channels are formed
on the lines where the Landau levels cross the Fermi level.
In more practical treatments in Refs.\cite{PhysRevB.46.4026, PhysRevB.52.R5535}, the QHECs are described as
``compressive" stripes separated by ``incompressive" insulating regions.
In the compressive stripes, the electrostatic potential is kept constant by the screening effect while the group velocity $\partial E_i/\hbar\partial k_i$
is finite. Therefore, eq.\eqref{eq_y-k_relation} does not hold inside the stripes and the wavenumber $k_i$ should also be kept constant.
In other words, eq.\eqref{eq_y-k_relation} only holds inside the incompressive regions.
As the value of $\varDelta y$, we should thus take the width of the incompressive stripes, which is generally much narrower than the compressive ones.
In Refs.\cite{PhysRevB.46.4026, PhysRevB.52.R5535}, such $\varDelta y$ is explicitly given for a simple classical electrostatic model of QHEC as
 \begin{equation}
  \varDelta y
\approx
  \sqrt{\frac{8|\varDelta E|\epsilon \epsilon_0}{\pi e^2(dn/dy)|_{y=y_i}}}\ ,
  \label{eq_dx}
 \end{equation}
where $i$ is the Landau index of the outer edge state, $\epsilon \epsilon_0$ is the dielectric permittivity of the matrix semiconductor,
$n(y)$ is the electron sheet density profile, $\varDelta E$ is the energy difference of the Landau levels.
The model has been used in analyzing many experimental works\cite{Oto2001,Masubuchi2006,PhysRevB.81.085329}.
$\varDelta E$ in the present case of exchange-aided Zeeman splitting can be written as $g\mu_{\rm B}B$,
where $g$ is the effective Land\'e g-factor, and $\mu_{\rm B}$ is the Bohr magneton.

Figure \ref{fig:phi-modulate}(b) shows an example of $V_{\rm C}$ dependence of $D$, calculated from eq.\eqref{eq_dx}
with the parameters noted in the caption.
These parameters are chosen to preserve semi-quantitative consistency with the analysis of the magnetic response described later.
To calculate $n(y)$ a function of $V_{\rm C}$ we employ ``frozen surface" model and the self-consistent Thomas-Fermi approximation in Ref.\cite{PhysRevB.52.R5535}.
Then $(dn/dy)|_{y=y_1}$ is obtained numerically from $n(y)$.
The characteristic behavior of the oscillation in Fig.\ref{fig:phi-modulate}(a) is qualitatively reproduced,
in that the oscillation phase more rapidly advances with negative $V_{\rm C}$ at lower $|V_{\rm C}|$.
The progress in the phase slows down, and the rewinding of the oscillation with increasing $|V_{\rm C}|$ begins
at the point indicated in the figure as ``Minimum point."
This behavior is qualitatively explained as follows  (a schematic of this description is in the supplemental material).
At low $V_{\rm C}$, the edge of 2DES lies near the end of the center gate, and the electrostatic confinement potential at the edge is soft,
leading to small $(dn/dy)|_{y=y_i}$ and large $\varDelta y$.
With increasing negative $V_{\rm C}$, the potential becomes steeper, lowering $\varDelta y$.
Then further increase in negative $V_{\rm C}$ causes softening of the potential and an increase in $\varDelta y$ again.
Since $\varDelta y$ must be smooth as a function of $V_{\rm C}$, $|d(\varDelta y)/dV_{\rm C}|$ decreases with negative $V_{\rm C}$, {\it i.e.} the oscillation period become slower, until the ``Minimum point'' roughly corresponding to the steepest edge confinement potential (maximum in $(dn/dy)|_{y=y_i}$),
and again increases with negative $V_{\rm C}$, resulting in the rewinding of the oscillation in $D$.

In spite of the obvious resemblance between Figs.\ref{fig:phi-modulate}(a) and (b),
quantitative fitting keeping the consistency with a response to the magnetic field is difficult.
We also tried the ``Fermi-level pinning" model for the surface states though it did not improve the quantitative agreement.
The discrepancy indicates the necessity to take into account the effects not considered, {\it e.g.} geometrical effect of the gate electrode.
However, the close resemblance between Fig.\ref{fig:phi-modulate}(a) and (b) still manifests the essential correctness of the scenario described so far.

 \begin{figure}[t]
  \includegraphics[width=1\linewidth]{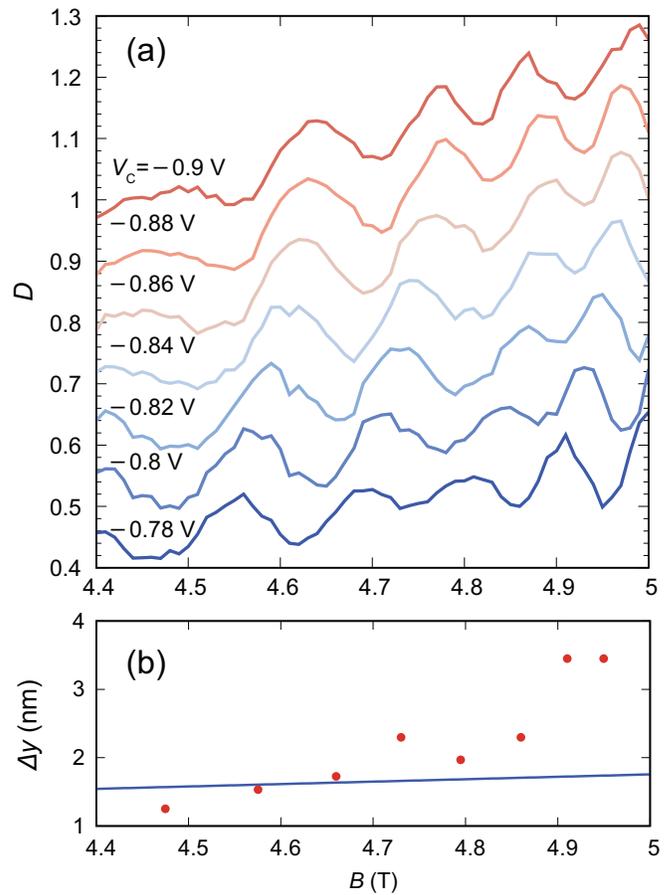}
  \caption{(a) Current partition rate $D$ as a function of $B$ within the plateau regime encompassing the filling factor $\nu=4$ for several values of Gate C voltage $V_{\rm C}$ (Traces are offset for clarity). $V_{\rm ac}=32.8~{\rm \mu V}-{\rm rms}$.
(b) $\varDelta y$ calculated from the equation $\varDelta y =(2/3)(h/e)/L\Delta B$ for $V_{\rm C}$=$-$0.86~V.
$\Delta B$ is given by twice the distance between adjacent oscillation peak and dip.
The blue line indicates $\varDelta y_{\rm average}=(\varDelta y (V_{\rm SL})L_{\rm SL}+\varDelta y (V_{\rm C})L_{\rm C}+\varDelta y (V_{\rm SR})L_{\rm SR}) /(L_{\rm SL}+L_{\rm C}+L_{\rm SR})$ from eq.\eqref{eq_dx} with the parameters used in Fig.\ref{fig:phi-modulate}(b).}
\label{fig:Bdep-Vdiff}
\end{figure}

As in eq.\eqref{eq_phi_AB}, the azimuth angle rotation is locked to the AB phase acquired from the magnetic flux piercing the incompressive regions.
This can be readily confirmed by the oscillatory behavior of $D$ versus $B$ shown in Fig. \ref{fig:Bdep-Vdiff}(a) with $V_{\rm C}$ as a parameter.
Because a single period $\varDelta B$ of the oscillation corresponds to 2$\pi$ rotation in $\phi$, $\varDelta y$ is given as $\varDelta y=(2/3)(h/e)/L\varDelta B$ from eq.\eqref{eq_phi_AB} in the local linear approximation of $B$-dependence of $\varDelta y$ in \eqref{eq_dx}, where $\varDelta E=g\mu_{\rm B}B$ (Details of this calculations are described in the supplemental material).
This gives $\varDelta y$ as a function of $B$ as shown in Fig.\ref{fig:Bdep-Vdiff}(b) for $V_{\rm C}$=$-$0.86~V.
The obtained values of $\varDelta y$ are much shorter than the magnetic length $l$=11~nm at $B$=5~T, being consistent with the view of compressive/incompressive stripes
while predicted $B$-dependence of $\varDelta y$ for constant $g$ against $B$ deviates from the experiment as indicated by Fig.\ref{fig:Bdep-Vdiff}(b).

\begin{figure}[t]
\includegraphics[width=1\linewidth]{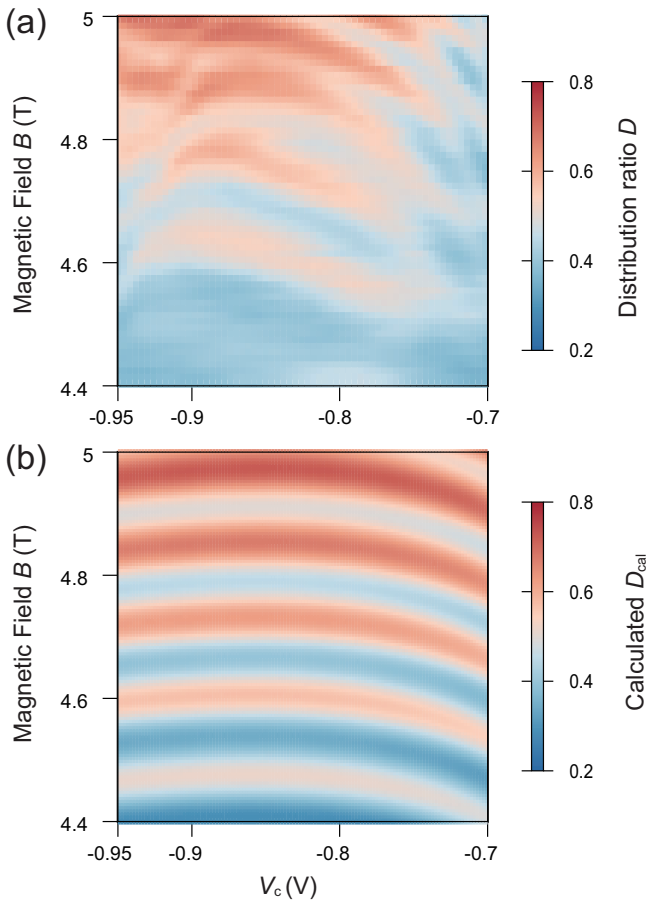}
\caption{(a) Color plot of the measured $D$ as a function of $B$ and $V_{\rm C}$. $V_{\rm SR}=V_{\rm SL}=-0.6V$. 
(b) Color plot of the theoretically given $D$ as a function of $B$ and $V_{\rm C}$.
$C_0=0.42B-1.45$, $C_1\sin(\theta)=0.1$ are used. Other parameters are the same as in Fig. 2.(b).}
\label{fig:Bdep}
\end{figure}

To see the overall tendency, measured and calculated values of $D$ are color plotted on $B$-$V_{\rm C}$ in Figs.\ref{fig:Bdep}(a) and (b) respectively.
The oscillation patterns appear as curved stripes in these plots. 
Such curving behavior is consistent with the interpretation of Fig.\ref{fig:phi-modulate} as follows.
With the increase of negative $V_{\rm C}$ from $-$0.7~V to about $-$0.8~V for fixed $B$, 
$(dn/dy)|_{y=y_1}$ increases, and thus $\phi$ decreases, corresponding to the up-going ridges.
After taking maximum at $V_{\rm C}\approx-0.86~{\rm V}$, 
$(dn/dy)|_{y=y_1}$ declines with the further increase in negative $V_{\rm C}$.

The observation of arc-like curving strongly supports the legitimacy of the analysis so far.
Similar arch-like curves were also observed at filling factor $\nu=2$ and $3$, but with smaller visibility.
At smaller filling factors, $B$ and hence $\varDelta y$ are larger, making $\theta$, the zenith angle, smaller as is discussed later.
As for the visibility, in Fig.\ref{fig:phi-modulate}, the oscillation under the present consideration is visible in the range of $V_{\rm C}$ from $-$0.7 to $-$1~V, and the visibility is the highest around the ``Minimum point." The tendency is common in the region of $B$ in Fig.\ref{fig:Bdep}, and the visibility does not change very much with $B$. Because an increase in $B$ means an enhancement in the rotation of $\phi$, the possibility of dephasing by numbers of $\phi$-rotation is eliminated.
Instead, we speculate that the simple model in Fig.\ref{fig-expl}(c) is approximately realized only around the steepest edge potential condition.
When the edge potential is soft, the QHECs have more chances to get the effect of local potential disorder.
As a result, the effective edge line fluctuates spatially creating local orbital angular momentum, which causes
inter-edge state scattering\cite{PhysRevB.45.11085}.
The above is a possible explanation of the dephasing in $\phi$-rotation.

From the above results and analysis, we can safely say that the current partition ratio $D$ reflects the azimuth angle rotation of FSQ traveling along the down edges of 
the gates. The rotation angle can be tuned via center gate voltage electrostatically. 

\section{Rotation in zenith angle}
\label{sec_zenith}
\begin{figure}[b]
\includegraphics[width=1\linewidth]{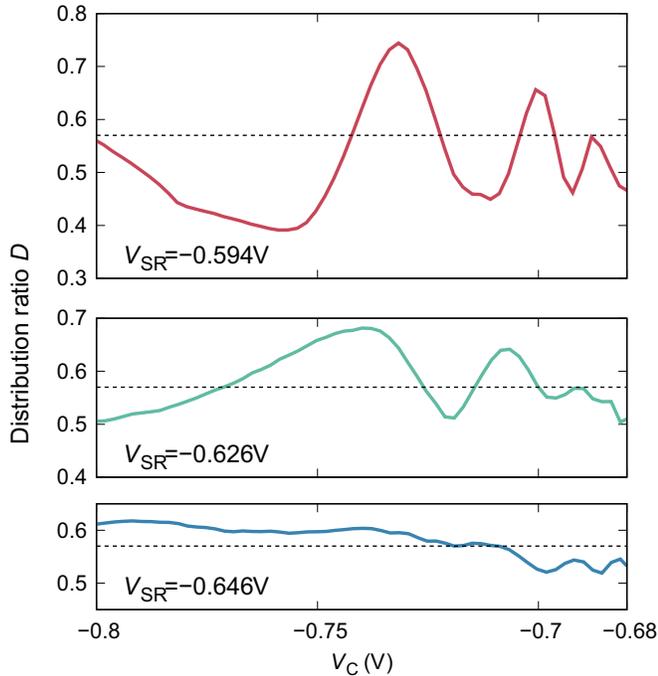}
\caption{Three oscillation patterns corresponding to three different values of $V_{\rm SR}$. $B=4.5~{\rm T}$ 
and $V_{\rm SL}=-0.605~{\rm V}$. $V_{\rm ac}=23.8~{\rm \mu V}$. 
The line for $D=0.57$ is indicated by broken lines 
to show that there is almost no movement in the oscillation baseline.
}
\label{fig:SideGateDiff}
\end{figure}
In Fig.\ref{fig:SideGateDiff}, we compare the oscillation patterns for three representative values of $V_{\rm SR}$, which strongly affects the
oscillation amplitude.
From $V_{\rm SR}=-0.594~{\rm V}$ the amplitude
 gradually decreases for further increase in negative $V_{\rm SR}$.
Characteristic features of the oscillation versus $V_{\rm C}$ so far observed do not change with
$V_{\rm SR}$ besides some phase shift is probably caused by a change in $\phi_0$. 
In the region $V_{\rm SR}>-0.594~{\rm V}$, the oscillation pattern changed drastically with a slight difference in $V_{\rm SR}$. This is probably because channel 2 gets into the spatial gap between Gate C and Gate SR. Hence,the region is out of present discussion.

\begin{figure}[t]
\includegraphics[width=\linewidth]{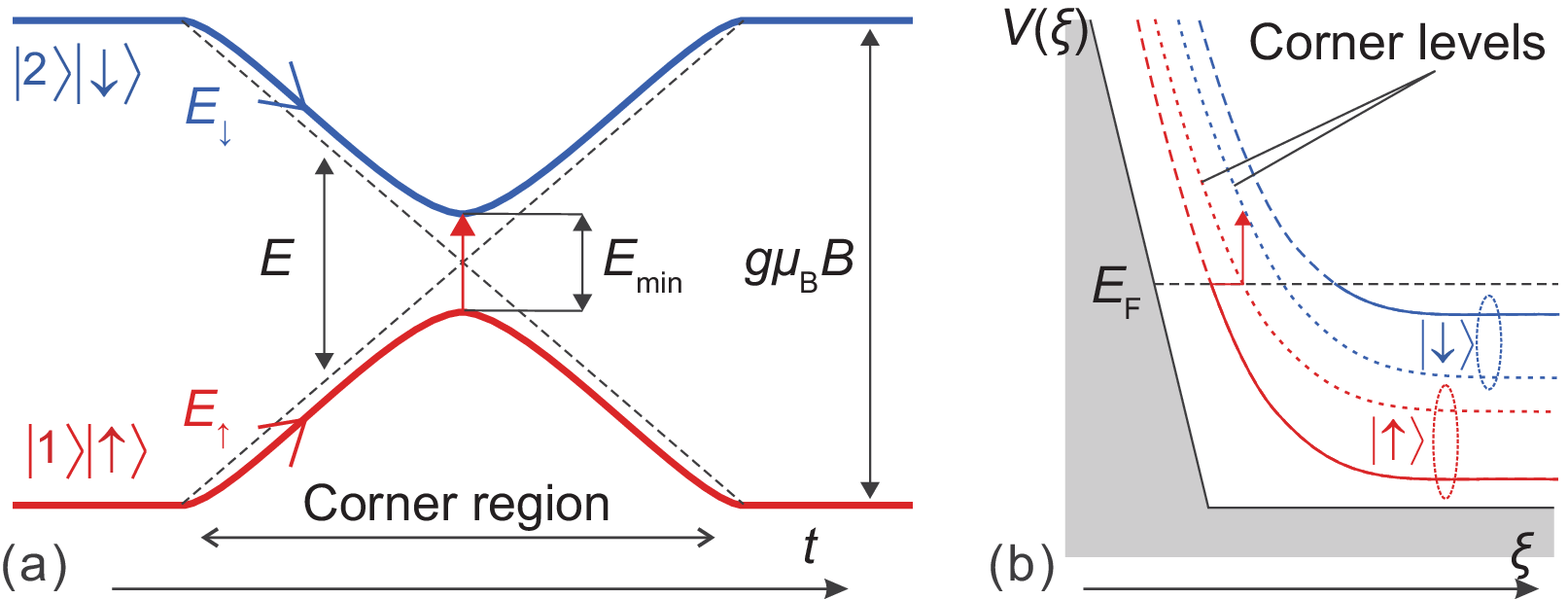}\\
\caption{(a) Schematic time evolution of quasi-eigenenergies $E_\uparrow=\bra{\uparrow}\bra{1}H_{\rm loc}(t)\ket{1}\ket{\uparrow}$
and $E_\downarrow=\bra{\downarrow}\bra{2}H_{\rm loc}(t)\ket{2}\ket{\downarrow}$.
(b) Illustration of spin-polarized edge states for a straight edge (solid and broken lines) and a corner (dotted lines). $\xi$ represents the distance from the
edge of infinite potential ($V(0)=\infty$).
For simplicity, the Landau levels are drawn in the single-electron picture.}
\label{fig:SideGatePotential}
\end{figure}

Hence we should look for the origin of the amplitude modulation in the zenith angle $\theta$,
which is determined when the wave packet turns the down-right corner of the Gate SR.
At the turning point, the time-dependent local Hamiltonian in \eqref{eq_wave_packet_travel} for the wave packet
should contain SOI terms: one from the in-plane potential gradient\cite{PhysRevB.76.153304};
the other from the Rashba and the Dresselhaus effects commonly observed in 2DES\cite{rolandwinkler2003}.
In the present case of spin-polarized QHEC, the former affects the effective Zeeman energy by spin-orbit effective field, while the latter kinematically rotates the spin.
Figure \ref{fig:SideGatePotential}(a) illustrates the time evolution of quasi-eigen energies for spin down and up, {\it i.e.} $E_\uparrow=\bra{\uparrow}\bra{1}H_{\rm loc}(t)\ket{1}\ket{\uparrow}$
and $E_\downarrow=\bra{\downarrow}\bra{2}H_{\rm loc}(t)\ket{2}\ket{\downarrow}$.
Around the center of the corner region $E=E_\downarrow-E_\uparrow$ takes the minimum value $E_{\rm min}$.

The transition in eq.\eqref{eq_Phi_SR} can then be taken as partially non-adiabatic tunneling.
By summarizing these effective time-localized SOI as $H_{\rm SOI}(t)$,
the probability $P$ of the inter-edge channel transition is given by slightly modifying a Landau-Zener type formula\cite{PhysRevA.23.3107} as,
\begin{align}
P\propto |\bra{\uparrow}\bra{1}H_{\rm SOI}\ket{2}\ket{\downarrow}|^2
\exp\left[-2\pi \frac{(E_{\rm min}/2)^2}{\hbar (dE/dt)}\right],
\label{Landau-zener}
\end{align}
where $dE/dt$ is the slew rate of $E$.
As indicated by arrows in Fig.\ref{fig:SideGatePotential}(b), the total process from $\ket{1}\ket{\uparrow}$ to $\ket{2}\ket{\downarrow}$
consists of non-adiabatic transition from $\ket{1}\ket{\uparrow}$ to $\ket{1}\ket{\downarrow}$ and adiabatic one from
$\ket{1}\ket{\downarrow}$ to $\ket{2}\ket{\downarrow}$.
The expression in \eqref{Landau-zener} tells that the slew rate as well as the minimum energy difference
strongly affect the transition probability.

A simple explanation of the tendency in Fig.\ref{fig:SideGateDiff} is led
from eq.\eqref{Landau-zener} and the electrostatic model in Refs.\cite{PhysRevB.46.4026,PhysRevB.52.R5535} as follows.
With increasing negative $V_{\rm SR}$, QHECs go away from the ``steepest potential" point, where the distance between the outer and the inner edges takes the minimum and
the radius of gyration $r_{\rm t}$ also is the shortest.
The smaller $r_{\rm t}$, the larger $dE/dt$ due to the shorter interaction time, and the potential gradient is larger, thus $E_{\rm min}$ is smaller.
From eq.\eqref{Landau-zener}, the transition probability $P$ takes maximum for the steepest edge potential condition.
As in Fig.\ref{fig:SideGateDiff}, this scenario tells that the steepest potential condition should correspond to $V_{\rm SR}>-$0.594~V,
which is considerably smaller than $-$0.8~V for $V_{\rm C}$.
The difference may come from the geometrical complexity in the real gate configuration.
As in the supplemental material, the equipotential lines around the corner change intricately, and the maximum of $P$ may appear
at smaller $V_{\rm SR}$ than the value at the steepest potential.

From the oscillation data in Fig. \ref{fig:SideGateDiff}, we can estimate the zenith angle $\theta$, assuming that the dephasing is ignorable at the largest amplitude region in $V_{\rm C}$ as follows.
Even if such ignorable dephasing is not the case, the lower limit of $\theta$ can be obtained from the analysis.
Let $t_{ij{\rm L}}$ be the complex transmission coefficients of the processes $\ket{i}\rightarrow\ket{j}$ at the down left corner of gate SL,
then from eq.\eqref{eq_Phi_SR}, the wave packet state that turns the corner and enters channel-1 to go to drain L is written as
\[
\ket{\Phi}_{\rm L}=\left(t_{\rm 11R}e^{ik_1L}t_{\rm 11L}+t_{\rm 12R}e^{ik_2L}t_{\rm 21L}\right)\ket{1}\ket{\uparrow}.
\]
For simplicity of expression, we write the complex transmission coefficients in the modulus-argument form as
$t_{\rm 11R}t_{\rm 11L}=t_1\cos(\theta/2)e^{i\varphi_1}$ and $t_{\rm 12R}t_{\rm 21L}=t_2\sin(\theta/2)e^{i\varphi_2}$.
This leads to the simple Young's double-slit result of the transmission coefficient $T_{\rm L}=\braket{\Phi|\Phi}_{\rm L}$
as
\begin{multline}
T_{\rm L}=t_1^2\cos^2(\theta/2)+t_2^2\sin^2(\theta/2)\\
+t_1t_2\sin\theta\cos(\phi+\varDelta \varphi).
\end{multline}
From the comparison with eq.\eqref{eq_D_C01},
\begin{equation}
C_0=t_1^2+(t_2^2-t_1^2)\sin^2({\theta}/{2}),\quad
C_1=t_1t_2.
\label{eq_osc_const}
\end{equation}

In Fig. \ref{fig:SideGateDiff}, the baseline of oscillation $C_0$ has almost no change while $C_1\sin\theta$ largely varies.
This fact tells from eq.\eqref{eq_osc_const} that $t_1$ and $t_2$ happened to be close to each other: $t_1\approx t_2$,
in the present condition (best visibility condition). Then $C_0\approx t_1^2\approx C_1\approx 0.57$.
In Fig. \ref{fig:SideGateDiff}, the largest amplitude gives $C_1\sin\theta$ as 0.17, which corresponds to $\theta \approx 17.4^{\circ}$.
This is the lower bound of the estimated $\theta$, which inevitably contains underestimation by dephasing.
To obtain a precise estimation of $\theta$, the oscillation of $C_1\sin\theta$ should be observed at least.
Unfortunately, in the present case, the maximum obtained value of $\theta$ is less than 90$^{\circ}$, and further analysis is difficult.
For more precise control of FSQ in the present scheme, the corner gates should be designed to create a sharper corner potential.
Furthermore, the dephasing should be reduced, {\it e.g.} by soft separation of the edges with an extra gate.

\section{Concluding remark}
We have studied the unitary operation of FSQs in QHECs with electric voltages on metallic gates.
This operation utilized the maximal entanglement between spin and edge channel orbitals.
The spin rotation in azimuth angle with voltage and with a magnetic field has been systematically studied.
A characteristic feature for spin appeared in the rotation in zenith angle, for which a new type of SOI
at a corner of edge channel was introduced and controlled with the gate voltage.
With the combination of these two
all electrical control of electron spin at spin-resolved quantum Hall edge states was achieved.

\begin{acknowledgments}
We thank Sathish Sugumaran and Leonid Beliaev for their collaboration 
in the initial stage of the present work.
This work was partly supported by Grants-in-Aid for Scientific 
Research on Innovative Areas, ``Nano Spin Conversion Science'' 
(Grant No. JP26103003), and by JSPS KAKENHI Grant Number JP19H00652.

\end{acknowledgments}

\bibliography{edgespin}

\end{document}